\documentclass[aps,nofootinbib,preprintnumbers,twocolumn,superscriptaddress]{revtex4-1}

\usepackage{hyperref}
\usepackage{amsmath,amssymb}
\usepackage{lmodern,bm}
\usepackage{multirow}
\usepackage{xspace}
\usepackage{colortbl}
\usepackage{tikz}
\usepackage{color}
\usepackage{mathrsfs}
\usepackage{wasysym}
\usepackage{yfonts}

\usepackage{slashed}

\usepackage{bbm}

\usepackage{fancyhdr}
\usepackage{mathbbol}
\usepackage{blindtext}

\def\bea{\begin{eqnarray}}
\def\eea{\end{eqnarray}}

\providecommand{\U}[1]{\protect\rule{.1in}{.1in}}
\newcommand{\gev}{\ensuremath{{\mathrm{\,Ge\kern -0.1em V}}}\xspace}
\newcommand{\mev}{\ensuremath{{\mathrm{\,Me\kern -0.1em V}}}\xspace}
\newcommand{\kev}{\ensuremath{{\mathrm{\,ke\kern -0.1em V}}}\xspace}
\newcommand{\fm}{\ensuremath{{\mathrm{\,fm}}}\xspace}

\usepackage{mfirstuc} 
\newcommand{\addReviewer}[2]{
  \expandafter\newcommand\csname #1\endcsname[1]{{\textbf{ \color{#2} \capitalisewords{#1}:\,##1}}}
  \expandafter\newcommand\csname #1cor\endcsname[2]{{\color{#2} \capitalisewords{#1}:\,\st{##1}{\textbf{##2}}}}
  \expandafter\newcommand\csname #1color\endcsname{#2}
  \expandafter\newcommand\csname #1todo\endcsname[1]{{\todo[inline,color=white!70!#2, caption={}]{\textbf{\capitalisewords{#1}}: ##1}}}
}

\usepackage{soul,color}
\definecolor{chromeyellow}{rgb}{1.0, 0.65, 0.0}

\addReviewer{ale}{chromeyellow}
\addReviewer{luc}{blue}
\addReviewer{angelo}{orange}

\newcommand{\be}{\begin{equation}}
\newcommand{\ee}{\end{equation}}

\begin{document}

\title{From the lineshape of the $X(3872)$ to its structure}

\newcommand{\infn}{INFN Sezione di Roma, P.le Aldo Moro 5, I-00185 Roma, Italy}
\newcommand{\catania}{INFN Sezione di Catania, V. Santa Sofia 64, I-95123 Catania, Italy}
\newcommand{\sap}{Dipartimento di Fisica, Sapienza Universit\`a di Roma, Piazzale Aldo Moro 5, I-00185 Roma, Italy}
\newcommand{\epfl}{Theoretical Particle Physics Laboratory (LPTP),Institute of Physics, EPFL, 1015 Lausanne, Switzerland}
\newcommand{\ias}{School of Natural Sciences, Institute for Advanced Study, Princeton, NJ 08540, USA}
\newcommand{\mift}{Dipartimento di Scienze Matematiche e Informatiche, Scienze Fisiche e Scienze della Terra,
Universit\`a degli Studi di Messina, Viale Ferdinando Stagno d'Alcontres 31, I-98166 Messina, Italy}

\author{Angelo~Esposito}
\affiliation{\epfl}
\affiliation{\ias}

\author{Luciano~Maiani}
\affiliation{\sap}
\affiliation{\infn}

\author{Alessandro~Pilloni}
\affiliation{\infn}
\affiliation{\mift}
\affiliation{\catania}

\author{Antonio~D.~Polosa}
\affiliation{\sap}
\affiliation{\infn}

\author{Ver\'{o}nica~Riquer}
\affiliation{\sap}

\begin{abstract}
From a study of the lineshape of the $X(3872)$, the LHCb collaboration measures a sizeable negative effective range. This cannot be reconciled with a shallow $D\bar D^*$ bound state hypothesis. Based on Weinberg's compositeness criterion, together with a theorem by Smorodinsky, it follows that the $X$ has to have a compact hidden charm structure
interacting with unbound $D\bar D^*$ pairs via short-distance color forces. This conclusion is strengthened by the general pattern recently emerging from exotic mesons. 
\end{abstract}

\maketitle

\section{Introduction}
The $X(3872)$ is the most studied among the many exotic hadrons which have been discovered since 2003~\cite{Chen:2016qju,Lebed:2016hpi,Esposito:2016noz,Guo:2017jvc,Olsen:2017bmm,Ali:2019roi,Brambilla:2019esw}. It is very often advocated that its mass being so close to the $D^0 \bar D^{*0}$ threshold\footnote{In the following, charge conjugation is understood.} is the proof of it being a loosely-bound molecule, the meson counterpart of the deuteron (`deuson'), with order few\kev's binding energy. This criterion is being  largely  used for other exotic states too, although none of them has been found as close to threshold as the $X$~\cite{Guo:2017jvc}.   
Nonetheless, the vicinity to threshold alone cannot be used as a criterion to tell a composite state of hadrons from a compact quark state. In the constituent diquark model~\cite{Maiani:2004vq}, for example, the vicinity to threshold is a purely accidental phenomenon. Indeed, recently, a good number of evidences seem to point towards a compact structure of the exotic states. These include high-multiplicity collisions~\cite{LHCb:2020sey}, and the recent discovery of four-quark states with hidden charm and strange valence quarks~\cite{BESIII:2020qkh,LHCb:2021uow}
and of four-charm resonances~\cite{LHCb:2020bwg}.

More refined analyses rely on the so-called Weinberg compositeness criterion~\cite{Weinberg:1965zz}, originally employed to investigate the possibility for the deuteron to be an elementary particle, rather than a composite bound state of proton and neutron. In a nonrelativistic theory, let $H_0$ be the free Hamiltonian describing states with the same quantum numbers as the deuteron. These are free $S$-wave proton and neutron pairs $|np(\bm{k})\rangle$, which form a continuum as a function of their relative momentum $\bm{k}$, and possibly an elementary deuteron, $|\mathfrak{d}\rangle$, of size comparable to the nucleons.

On the one hand, if such an elementary deuteron is absent from the theory,  strong interactions may nevertheless generate a molecular deuteron, $|d\rangle$, with some binding energy $-B$, as a result of the interaction between $n$ and $p$ due to a potential $V$. 
On the other hand, if the eigenstates of $H_0$ feature the elementary $|\mathfrak{d}\rangle$, the interaction potential $V$ will now also contain pointlike interactions between $\mathfrak{d}$ and the $np$ pairs. The physical state $|d\rangle$ will be an eigenstate of $H_0+V$ arising from all these interactions. The overlap of the interacting deuteron $|d\rangle$ with the elementary one is measured by $Z\equiv {|\langle\mathfrak{d}|d\rangle|}^2$. 
The deuteron state (normalized to unity) can then be expressed in terms of the eigenstates of $H_0$ as $|d\rangle= \sqrt{Z} |\mathfrak{d}\rangle +\int \frac{d^3 {\bm k}}{(2\pi)^3} \, C(\bm k) |np(\bm k)\rangle$, with $\int \frac{d^3{\bm k}}{(2\pi)^3}\, |C(\bm k)|^2=1-Z$.
When $Z=0$ no elementary deuteron is present in the theory, while when $Z=1$ the theory is free and no $|np\rangle $ component is present in $|d\rangle$. 
$Z>0$ implies that an elementary $|\mathfrak{d}\rangle$ does exist, and that the so-called effective range, $r_0$, is negative, a condition that, as we explain in detail below,  is incompatible with a molecular shallow bound state.
The physical deuteron results from the interaction of the elementary one with the $|np\rangle$ pairs. In this case, the existence of a bound state generated solely by 
$np$ interactions is not necessary to explain the dynamics of the system.
 
The same argument can be cast in a nonrelativistic field theory (see, e.g.,~\cite{Chen:2013upa,Guo:2017jvc}). Consider a canonical field, $\Phi(x)$, creating the elementary deuteron, $\Phi(0)|0\rangle = |\mathfrak{d}\rangle$, and such that $\langle d | \Phi(0) |0\rangle = \sqrt{Z}$.
When $Z$ is different from zero, the propagator for $\Phi$ features a pole in correspondence of the mass of the deuteron, with residue $Z$. From the K\"all\'en-Lehmann representation one can indeed show that $0\leq Z\leq 1$.\footnote{The Lehmann normalization condition is  $Z+\int_0^\infty d\mu \, \sigma(\mu)=1$, where $\sigma(\mu)\geq 0$ is the contribution of multiparticle states in the spectral representation of the complete propagator and $Z$ is the residue at the one-particle pole.} As said, the case $Z=1$ corresponds to a free elementary particle, for which $\Phi(0)|0\rangle$ does not couple to multiparticle states.
The case $Z=0$ can be interpreted as the condition for the particle to be composite, since its field does not appear in the Lagrangian: the state is generated dynamically, and the propagator is saturated by multiparticle contributions. 
This reasoning is compatible with the statements written in the original Weinberg's paper~\cite{Weinberg:1965zz}, where it is observed that ``{\it an elementary deuteron would have $0<Z<1$}"
~\footnote{
Although present in the literature, we think that the interpretation of $Z$
as the mixing probabilty between a molecule and a compact state is a too naive interpretation of Weinberg's formalism. $Z$
measures the projection of the $X$ state on the discrete part of the eigenstates
of $H_0$, which is different from the basis of the
physical states. This is indicated by the fact that $Z = 1$ 
~represents a free state decoupled from $D^* \bar D$, certainly not the case of any physical state with the same quantum numbers as $D^*\bar D$.
Rather, $0 < Z < 1$ indicates that $D^* \bar D$ states are an incomplete
set, in the space of the eigenstates of $H_0$, and they have to be supplemented
by states belonging to the discrete spectrum of $H_0$: bare compact states do
exist. Also, $0 < Z < 1$ does not say anything about the
existence of bound states in the inter-hadron $D^* \bar D$ potential,
i.e. molecules: the interaction could be driven by the compact state only, which would be consistent with no bound molecule at all. 
 }.

A key quantity in the study of composite states is the effective range, defined through the low energy expansion of the scattering phase. For a shallow molecule in a purely attractive potential, $r_0$ is positive, of the order of the range of the potential~\cite{smorodinsky,LandauSmorodinsky}. Indeed, for the deuteron one has $r_0 \simeq + 1.7\fm$, in nice agreement with $1/m_\pi\simeq1.4\fm$. A value of the effective range that is negative and well beyond the range of the potential is  a sign that the dynamics of the system cannot be explained without the presence of a compact, elementary state. Indeed, Weinberg  notes that {\it ``the true token that the deuteron is composite is an effective range $r_0$ small and positive rather than large and negative"}~\cite{Weinberg:1965zz}.

As we saw, the deuteron nicely checks all the requirements to be a loosely-bound molecule. But what about the $X(3872)$? The LHCb collaboration recently published a high-statistics analysis of its lineshape~\cite{Aaij:2020qga}. Since the fit parameters were highly correlated, the error analysis is particularly delicate, and we focus first on best fit values neglecting error correlations. The result is a value of $Z\simeq 0.14$ and in a sizeable negative effective range $r_0\simeq -5.34\fm$

After this work appeared in preprint form,  the error correlations of LHCb data  have been analysed in~\cite{Baru:2021ldu}. 
As a consequence,  we obtain $r_0$ in the range
\be
-2.0 ~{\rm fm}> r_0 > -5.3~ {\rm fm} \label{rangefin}
\ee

If this is confirmed, it entails a clear conclusion: the $X$ is generated by an elementary core with a sizeable dressing of $D\bar D^*$ states in the continuum, due to their short-distance QCD coupling. 
This is qualitatively different from a bona fide loosely-bound molecule, for which $Z=0$ and $r_0>0$.

This same conclusion was suggested from studies on the production of the $X(3872)$~\cite{Bignamini:2009sk,Esposito:2017qef} although different conclusions are reached in~\cite{Artoisenet:2009wk}. The comparison to deuteron, has been further 
studied in~\cite{Esposito:2020ywk}: it is observed that the production of a molecule in high-multiplicity final states appears to be qualitatively different from that observed for the $X(3872)$ in the same conditions.   

In this work we first review the compositeness criteria, unifying different treatments. In light of this, we then re-examine the recent LHCb data, to extract the values of $Z$ and $r_0$.

\section{Scattering amplitude and compositeness}
Analogously to~\cite{Weinberg:1965zz}, call $H_0$ the Hamiltonian representing the QCD quark interaction generating compact color singlets. In the open charm sector, the spectrum of $H_0$ contains free $D$, $D^*$ mesons and their antiparticles. In the hidden charm $J^{PC} = 1^{++}$ sector, it contains free charmonia, and possibly hidden charm tetraquarks. The potential $V$ is, instead, responsible for the $S$-wave $DD^*$ interactions, for example mediated by pion exchange. In presence of a compact tetraquark, $V$ is also responsible for its trilinear interaction to the $D\bar D^*$ continuum (as well as to other pairs of mesons).

Consider the low-energy $D\bar D^*$ scattering in their center-of-mass system. At low relative momenta, the general expression for the $S$-wave scattering amplitude is
\begin{align} \label{generalf}
f = \frac{1}{k \cot \delta(k) - i k} = \frac{1}{-\kappa_0 + \frac{1}{2} r_0 k^2 + \dots- i k }\,,
\end{align}
where $\delta(k)$ is the scattering phase and $\kappa_0 = a_0^{-1}$ is the inverse of the scattering length. Dots represent higher order terms in the expansion of the scattering phase. 

In presence of a state $X$ below threshold by an energy $B$, the amplitude will have a pole for $-ik = \sqrt{2\mu B}\equiv \kappa$, with $\mu$ the reduced $D\bar D^*$ mass. This implies the relation
\begin{align} \label{eq:kappa}
\kappa = \kappa_0 + \frac{1}{2} r_0 \kappa^2\,,
\end{align}
that must be satisfied regardless of the nature of the $X$. 
A discrimination between different structures of the $X$ comes from the effective range. 

In Weinberg's treatment one defines a coupling constant, $g=\langle D \bar D^*|V|X\rangle$, which, through the completeness relation for the eigenstates of $H_0$ can be related to $B$ and $Z$ as\footnote{The expression~\eqref{eq:gWeinberg} assumes the more common convention for which the volume element in momentum space is $d^3{\bm k}/(2\pi)^3$, rather than just $d^3{\bm k}$, as done in~\cite{Weinberg:1965zz}.}
\begin{align} \label{eq:gWeinberg}
g^2 = \frac{2\pi \kappa}{\mu^2}(1-Z)\,.
\end{align}
Iterating the pole amplitude, one obtains a geometrical series which sums up to the scattering amplitude to order $k^2$~\cite{Ali:2019roi}. Comparing with the formula for the scattering amplitude to this order, 
Weinberg~\cite{Weinberg:1965zz} obtains
\begin{subequations}
\begin{align} 
\kappa_0^{-1}&=2\frac{1-Z}{2-Z} \kappa^{-1} + O (1/m_\pi)\,, \label{kappa0} \\
r_0&=-\frac{Z}{1-Z}\kappa^{-1} + O (1/m_\pi)\,,  \label{r02}
\end{align}
\end{subequations}
where $1/m_\pi$ represents the typical range of the interaction between $D$ and  $\bar D^*$, encoding the non-universal corrections to the shallow bound state limit. We note that the expressions above indeed satisfy the relation~\eqref{eq:kappa} for any value of $Z$.

From Eq.~\eqref{r02} it is clear that for an elementary $X$ with $Z>0$, the effective range must be negative, $r_0<0$. For a genuine loosely-bound state ($Z=0$), instead, $r_0 = O(1/m_\pi)$. As we show in the next section, this is not the end of the story. In this case, for purely attractive binding force, the sign of the non-universal part of the effective range is  fixed to be positive.

Landau~\cite{landaup} makes a similar analysis for the scattering in presence of a state below threshold, starting from the interaction Lagrangian
$\mathcal{L}_I = g_\text{L} X D \bar D^*$.\footnote{Landau's and Weinberg's couplings have different normalizations, and are related by $g_\text{L}^2 = 8 \mu m_X^2 g^2$. }
The scattering amplitude can be written as~\cite{Polosa:2015tra}
\begin{align}
\begin{split} \label{pole2}
f&=-\frac{1}{8\pi m_X}\,g_\text{L}^2\,\frac{1}{(p_D+p_{D^*})^2-m_X^2} \\
&\simeq -\frac{1}{16\pi m_X^2}\,g_\text{L}^2\,\frac{1}{B+E}\,,
\end{split}
\end{align}
where $m_X=m_{D^0}+m_{\bar{D}^{*0}}-B$ is the mass of the $X$ and $E=k^2/2\mu$ is the center-of-mass energy of the $D\bar D^*$ pair.
On the other hand, from Eq.~\eqref{generalf}, near the pole we may write
\begin{align} \label{pole3}
f \simeq -\frac{1}{ \kappa^2 + k^2} \frac{2\kappa}{1-r_0\kappa}=-\frac{\kappa}{\mu(1-r_0\kappa)} \frac{1}{E+B}\,.
\end{align}
Comparing with \eqref{pole2}, one finds
\begin{align} \label{landg}
g_\text{L}^2=16\pi m_X^2\frac{\kappa}{\mu(1-r_0\kappa)} \,,
\end{align}
 This is consistent with Weinberg's result~\eqref{eq:gWeinberg} with the identification
 \be
 Z=\frac{-r_0\kappa}{1-r_0\kappa} \label{ltow}
 \ee

The result \eqref{landg} for the coupling can be used to compute the $X\to D\bar{D} \pi $ branching ratio as done in~\cite{Polosa:2015tra}.
Since $r_0$ is multiplied by $\kappa$, $g_\text{L}$  is not sensitive to the value of the effective range, for very small binding energies.

\section{The effective range of a molecule} \label{effrad}

As anticipated, Landau and Smorodinsky add a specific consideration regarding the value of $r_0$ for a molecule, constraining the terms of order $1/m_\pi$, left unspecified in Eq.~\eqref{r02}~\cite{LandauSmorodinsky,Blatt:1952ije}, when $Z=0$. Their conclusion extends what happens for the deuteron to a general theorem:  {\it shallow bound states with  purely attractive binding force give always $r_0> 0$}. 
For convenience of the reader, we report a proof of the theorem, following Bethe's derivation~\cite{Bethe1949}.

Consider the Schr\"odinger's equation for the radial wave function of the molecular constituents,
\begin{align} \label{pot}
u_k^{\prime\prime}(r) +\big[k^2-U(r)\big]u_k(r)=0\,,
\end{align}
with $U(r) \equiv 2 \mu V(r)$, {$V(r) < 0$} being the potential, which is assumed to be attractive everywhere. 
We consider the wave function for two values of the momentum: $u_{k_{1,2}}\equiv u_{1,2}$. A simple manipulation leads to the identity
\begin{align} \label{eqrad}
u_2u_1^\prime - u_2^\prime u_1 \Big |_0^R=(k_2^2- k_1^2)\int_0^R dr \, u_2 u_1\,,
\end{align}
with $R$ fixed and much larger than the range of the potential, $R\gg 1/m_\pi$.

Consider now the free equation, $\psi_k^{\prime\prime}(r) +k^2\psi_k(r)=0$, from which we also obtain
\begin{align} \label{eqpsi}
\psi_2\psi_1^\prime - \psi_2^\prime \psi_1 \Big |_0^R=(k_2^2- k_1^2)\int_0^R dr \, \psi_2 \psi_1\,.
\end{align}
Normalizing to unity at $r=0$, the general expression for $\psi_k$ is
\begin{align}
\psi_k(r) = \frac{\sin(kr+\delta(k))}{\sin\delta(k)}\,,
\end{align}
so that $\psi^\prime_k(0) = k \cot \delta(k)$.
The radial wave function $u_k$ vanishes at $r=0$, and we normalize it so that it tends exactly to the corresponding $\psi_k$ for large enough radii. 

With this proviso, we subtract~\eqref{eqrad} from~\eqref{eqpsi}, to obtain
\begin{align} 
\begin{split} \label{phase2}
&k_2\cot \delta(k_2)-k_1\cot \delta(k_1)  \\ 
&\qquad =(k_2^2- k_1^2)\int_0^\infty dr \, (\psi_2 \psi_1 - u_2 u_1)\,.
\end{split}
\end{align}
Here we used the fact that, for $R$ large enough $u_k\to \psi_k$, as well as the fact that $u_k(0)=0$. Moreover, we have extended the integral to infinity, given that it is now convergent due to the same  asymptotic behavior of $\psi_k$ and $u_k$.

We are now ready to compare the result with the parameters of the scattering amplitude~\eqref{generalf}. First we set $k_1=0$ which, recalling that $\lim_{k_1\to0} k_1 \cot \delta(k_1) = -\kappa_0$, gives
\begin{align}
k_2 \cot\delta(k_2) = - \kappa_0 + k_2^2 \int_0^\infty dr\, \left(\psi_2 \psi_0 - u_2 u_0\right)\,.
\end{align}
Finally, for a shallow bound state, one can further expand for small momenta, $k_2 \cot \delta(k_2) = - \kappa_0 + \frac{1}{2}r_0 k_2^2 + \dots$, thus finding
\begin{align}
r_0=2\int_0^\infty dr\, (\psi _0^2-u_0^2)\,.
\end{align}
Since $u_0(0) = 0$ and $\psi_0(0) = 1$ and both go to the same limit at infinity, calling $\Delta(r) = \psi_0(r) - u_0(r)$, we have $\Delta(0) = +1$ and $\Delta(\infty) = 0$. Moreover, the equations of motion imply $\Delta^{\prime\prime}(r) = -U(r)u_0(r)>0$ for an attractive potential. In presence of a single bound state, where $u_0(r)$ does not have nodes, we get $\Delta''(r) > 0$ everywhere, hence proving that $\psi_0(r) > u_0(r)$.
Given that $u_0 < \psi_0$ everywhere, this proves that $r_0 > 0$ for a shallow bound state.

This is the result quoted in~\cite{LandauSmorodinsky} where the further approximation $\psi_0=1$ is made, consistently with the fact that  $\psi_0\simeq 1-r/a_0\simeq 1$ for a shallow bound state and $r$ within the range of the potential.

\section{A new look at  the LHCb data}

The discussion so far refers to the elastic scattering of two (stable) particles that resonate on a shallow level. This is not the case for the $X(3872)$, that couples also to other channels (most notably $J/\psi\pi\pi$, where it is most easily detected). These arguments can still approximately hold, provided that a single channel dominates the $X(3872)$, as might be suggested by the seemingly large branching ratio for the $X\to D\bar D \pi$ decay~\cite{Li:2019kpj}. Moreover, the finite width of the $D^{*}$ must be taken into account.
Possible extensions have been proposed in~\cite{Baru:2003qq,Sekihara:2014kya,Guo:2015daa,Oller:2017alp,Matuschek:2020gqe}. Anyway, the formulae require some modifications to be suitable for data analysis. As an introduction, we first make explicit the connection between the formula for the scattering  amplitude, Eq.~\eqref{generalf}, and  the usual nonrelativistic Breit-Wigner formula, writing
 \begin{align} \label{nrbw}
f=-\frac{\frac{1}{2}g^2_\text{BW}}{ E - m_\text{BW} + \frac{i}{2} g^2_\text{BW} k}\,,
\end{align}
with $g^2_\text{BW} = -2/\mu r_0$ and $m_\text{BW} = \kappa_0 /\mu r_0$. For $r_0 < 0$ and $m_\text{BW} \gg \mu g_\text{BW}^4$, this expression describes an ordinary resonance above threshold, having width $\Gamma \simeq g^2_\text{BW} \sqrt{2 \mu m_\text{BW}}$. 
Recall, however, that we are interested in a shallow state below threshold. 

Recently, LHCb studied the lineshape of the $X(3872)$ with a high statistics sample~\cite{Aaij:2020qga}. Two main models were considered. The first one is a standard $S$-wave relativistic Breit-Wigner, which yields a width $\Gamma_X \simeq 1.2\mev$. However, this model is realistic only if the coupling of $X\to D^0 \bar D^{*0}$ is not very large---which does not seem to be the case. 
The second model is based on the Flatt\'e amplitude~\cite{Flatte:1976xu,Baru:2004xg}, as extended in~\cite{Hanhart:2007yq}. Since we are interested in the sign of the effective range, fitting a larger number of parameterizations would be ideal to reduce model bias~\cite{Fernandez-Ramirez:2019koa}. For the time being, we focus on the published Flatt\'e one.
Indicating with $N$ an unknown normalization constant, $m^0_X$ a bare parameter that controls the mass of the $X$, $\delta = m_{D^{*-}} + m_{D^{+}} - m_{\bar{D}^{*0}} - m_{D^0} = 8.2\mev$ the isospin splitting, and $\mu$ and $\mu_+$ the reduced masses of the neutral and charged $D\bar D^*$ pairs, we write
\begin{widetext}
\begin{align} \label{flattelhcb}
    f\left(X\to J/\psi \pi^+\pi^-\right) = -\frac{N}{E - m^0_X + \frac{i}{2} g_\text{LHCb} \left(\sqrt{2 \mu E} + \sqrt{2 \mu_+ (E - \delta)}\right) + \frac{i}{2} \left(\Gamma^0_\rho(E) + \Gamma^0_\omega(E) + \Gamma^0_0\right) } \,,
\end{align}
\end{widetext}

The information about other decay channels is encoded in $\Gamma^0_{V}(E)$---the running bare widths of $X \to J/\psi \,V$---and $\Gamma^0_0$---a constant that takes into account further possible channels. We use ``bare'' to stress that these numbers are not the mass and partial widths of the $X$ yet, but are connected to them. We remind the reader that $E$ is the kinetic energy, related to the $J/\psi \pi^+ \pi^-$ invariant mass by $m_{J/\psi\pi^+\pi^-} = E + m_{D^0} + m_{\bar{D}^{*0}}$.

 In order to match with the single channel analysis of the previous Section, we can set $\Gamma^0_\rho = \Gamma^0_\omega = \Gamma^0_0 = 0$. 
The amplitude now describes a stable $X(3872)$ that couples to $D^0 \bar D^{*0}$ and $D^+  D^{*-}$ only. Since the charged threshold is further away, we can expand the amplitude for small kinetic energies, and get
\begin{widetext}
\begin{equation}
    f\left(X\to J/\psi \pi^+\pi^-\right) = -\frac{N \frac{2}{g_\text{LHCb}}}{\frac{2}{g_\text{LHCb}}(E - m^0_X) - \sqrt{2\mu_+\delta} + E \sqrt{\frac{\mu_+}{2\delta}} + ik}
    \label{flattenew}
\end{equation}
\end{widetext}
In their analysis, the LHCb collaboration fixes $m_X^0 = - 7.18$~MeV (with respect to the $D^0\bar D^{*0}$ threshold) and fits $g_\text{LHCb} = 0.108\pm 0.003$. From this, we obtain the inverse scattering length and the effective range:
\begin{subequations}
\begin{align}
\kappa_0 &= -\frac{2m_X^0}{g_\text{LHCb}} - \sqrt{2\mu_+\delta} \simeq 6.92~\mev\,,
\label{resultsa} \\
r_0 &= -\frac{2}{\mu g_\text{LHCb}} - \sqrt{\frac{\mu_+}{2\mu^2\delta}} \simeq -5.34~\fm\,.
\label{resultsb}
\end{align}
\end{subequations}
To derive $\kappa$, we use the consistency condition Eq.~\eqref{eq:kappa}.
The physical root reduces to $\kappa_0$ for $r_0\to 0${, thus}
\begin{align}
\kappa^{-1}\simeq33~\fm\,, \quad \text{ and } \quad B =\frac{\kappa^2}{2\mu} \simeq 18~\kev\,.
 \label{binden}
\end{align}

According to Ref.~\cite{Baru:2021ldu}, LHCb data give a relatively good determination of the ratio $m^0_X/g_\text{LHCb}$, while $g_\text{LHCb}$ is poorly determined: it has a very shallow $\chi^2$ distribution that allows values of $g_\text{LHCb}$ even ten times larger than the best fit value used in Eq.~\eqref{resultsb}. Neglecting the first term in~\eqref{resultsb}, one obtains the upper bound to $r_0$ anticipated in Eq.~\eqref{rangefin}.

The range of $r_0$ in Eq.~\eqref{rangefin} is definitely negative. Incidentally, qualitatively similar results were already found in~\cite{Meng:2013gga,Kalashnikova:2009gt,*Kalashnikova:2018vkv} based on older datasets.
This is an unequivocal evidence that the dynamics of the system is driven by the hard compact structure of  the $X$. 
No other conclusion about the nature of the $X(3872)$ can be drawn.
The situation is similar to the large $p_T$ production of the $X$ in hadronic collisions~\cite{Esposito:2015fsa,Meng:2013gga}, where a hard structure is definitely required, without necessarily implying the presence of a molecular one.
From Eqs.~\eqref{eq:gWeinberg}, \eqref{r02}, and~\eqref{resultsb}, one also gets
\begin{equation}
 g^2 = \frac{2\pi \kappa}{\mu^2} \frac{g_\text{LHCb}}{g_\text{LHCb}\left(1 + \frac{\kappa}{\mu}\sqrt{\frac{\mu_+}{2\delta}}\right)+ \frac{2\kappa}{\mu}} \,,
\end{equation}
which explicits the relation between the Weinberg and LHCb couplings. 

Using \eqref{ltow} with \eqref{rangefin} and the value of $\kappa$ in \eqref{binden}, one finds
\be
0.052 < Z < 0.14
\ee


\section{Conclusions}

Since the first observation of the $X(3872)$, the debate regarding its nature (and consequently that of other exotic states too) has been intense. To discriminate between a compact tetraquark nature and a loosely-bound molecular one, it is crucial to find quantities that are qualitatively different in the two instances.

In this work we show that, combining the arguments by Weinberg~\cite{Weinberg:1965zz} and by Landau and Smorodinsky~\cite{LandauSmorodinsky,smorodinsky,Blatt:1952ije,Bethe1949}, one can identify one such quantity in the effective range. If the dynamics of the $X$ is that of a purely shallow molecule, the effective range must be strictly positive and of the order of the inverse pion mass, as it happens for the deuteron. If, instead, the dynamics is that of an elementary object, the effective range is negative and in magnitude much larger that the inverse pion mass. 

The most recent LHCb analysis shows that the latter condition is met by experimental data, at least for the central values. If confirmed beyond experimental errors,  the effective range value could represent the smoking gun demonstrating that the $X(3872)$ has an elementary core which interacts with the unbound $D\bar D^*$ pair through short-distance QCD. This does not specify {\it a priori} whether it is a charmonium or a hidden-charm tetraquark, although the latter seems more natural, looking at the other exotic candidates in the same mass region.

This conclusion is strengthened by the general pattern emerging with the straightforward interpretation~\cite{Becchi:2020uvq} of the di-$J/\psi$ $X(6900)$ resonance~\cite{LHCb:2020bwg}, and the discovery of  $Z_{cs}(3985)$ and $Z_{cs}(4003)$~\cite{BESIII:2020qkh,LHCb:2021uow}, reproducing in the strange quark sector the situation observed with the $X(3872)$ and the $Z_c(3900)$~\cite{Maiani:2021tri}. 

The recent observation of a doubly charm state~\cite{LHCb:2021vvq} 
 (predicted, among others, in~\cite{DelFabbro:2004ta,Carames:2011zz,Dias:2011mi,Hyodo:2012pm,Esposito:2013fma}) 
 will make it possible an independent study of its lineshape, along the lines shown here~\cite{LHCb:2021auc,*Du:2021zzh}.

\begin{acknowledgments}

A.D.P. wishes to thank Mikhail Mikhasenko for providing a copy of the original Smorodinsky's paper. A.E. is grateful to Riccardo~Rattazzi and Andrea~Wulzer for illuminating discussions on the nonrelativistic theory of compact tetraquarks.
The work by A.E. is supported by the Swiss National Science Foundation under contract 200020-169696, and through the National Center of Competence in Research SwissMAP, has received funding from the European Union's Horizon 2020 research and innovation programme under the Marie Sk{\l}odowska-Curie grant agreement No.~754496.
\end{acknowledgments}

\appendix

\medskip

\bibliographystyle{apsrev4-1.bst}
\bibliography{biblio}

\end{document}